\documentclass[aps,pra,showpacs]{revtex4}
\usepackage{graphicx}
\usepackage{amsmath,amssymb,amsthm,bbm,latexsym}
\usepackage{amsthm}
\usepackage{amsbsy}

\begin{document}

\title{Optimal Broadcasting of Mixed States}
\author{Gui-Fang Dang, Heng Fan}
\affiliation{Institute of Physics, Chinese Academy of Sciences,
Beijing 100080, China.} \pacs{}

\begin{abstract}
The $N$ to $M$ ($M\ge N$) universal quantum broadcasting of mixed
states $\rho^{\otimes N}$ are proposed for qubits system. The
broadcasting of mixed states is universal and optimal in the sense
that the shrinking factor is independent of input state and achieves
the upper bound. The quantum broadcasting of mixed qubits is a
generalization of the universal quantum cloning machine for
identical pure input states. A new pure state decompositions of the
identical mixed qubits $\rho^{\otimes N} $ are obtained.
\end{abstract}

\maketitle

\section{\protect\bigskip Introduction}

An unknown quantum state can not be cloned perfectly, i.e. the no-cloning
theorem\cite{Wootters,Barnum}, which is one fundamental theorem of quantum
mechanics and quantum information. However, it does not mean that we cannot
clone quantum states imperfectly or probabilistically. So far, much effort
has been devoted into imperfect cloning and probabilistical cloning of
quantum states\cite{Buzek}-\cite{Fan2}.

The quantum cloning of pure states has already been well studied,
for reviews, see Refs.\cite{SIGA,CF,F00}. However, the study of
cloning of identical
mixed states has only recently attracted some attentions
\cite{Cirac,Giacomo,Buscemi,FLS}. In a review paper in Rev. Mod. Phys. \cite{SIGA}%
, the first open question raised by the authors is the quantum
cloning of mixed states. In this paper we will study this problem
with input being the identical mixed qubits. To avoid confusion, we
use "broadcasting" instead of "cloning" for copying of mixed states
since the output state of the processing is in general not
factorized.

For pure state input without a prior information, we can construct the
universal quantum cloning machine (UQCM), i.e., the quality of its output
does not depend on the input. We can use distance like parameters between
input and output to quantify the merit of the quantum cloning machine. For
example, the Hilbert-Schmidt norm was used in Ref.\cite{Buzek} and later the
fidelity is well accepted for various quantum cloning machines \cite%
{Bruss,Werner,SIGA}. The quantification of the merit of the quantum
broadcasting of mixed states seems more complicated. It is recently
pointed out that the mixed state input cannot be universally
broadcasted if we use the fidelity to describe
the merit of the broadcasting \cite{CC}. It is shown that for $N$ to $M$ (%
$N\geq 2$) quantum broadcasting, where $N$ identical input states are
quantum broadcasted to $%
M $ states, the fidelity between single qubit mixed input and output
is input dependent unless the input is pure or completely mixed.
However, the shrinking factor was used as the parameter to quantify
the merit of the broadcasting of identical mixed qubits \cite{FLS}.
In this paper, for the reasons as the following, we will still use
the shrinking factor to quantify how well the mixed qubits are
copied. For \textit{universal} quantum broadcasting, the single
qubit output state generally
should satisfy the scalar relation $\rho _{out}^{single}=\eta \rho + \frac{%
1-\eta }{2}I$ which will appear later (\ref{9}), where $\rho $ is
one of the identical mixed input qubits, $I$ is the identity
operator in two-dimensional space, and $\eta $ is what we called
shrinking factor. We can find that the output state is actually a
mixed state constructed by the original input with
probability $\eta $ and the completely mixed state $I/2$ with probability $%
1-\eta $. In quantum information processing, it is well accepted that the
completely mixed state contains no information. Thus all information of the
copies $\rho _{out}^{single}$ is in state $\rho $. And we then assume that
the shrinking factor $\eta $ can be used as one parameter of the merit of
the broadcasting of mixed states. It can also be understood as that the quantum state $%
\rho _{out}^{single}$ is the output of a quantum depolarization channel with
input $\rho $, where $1-\eta $ is related with the noise of this channel.

The optimal fidelity of $N$ to $M$ UQCM for pure state has already
been obtained in Refs.\cite{Gisin,Bruss,Werner}. The optimal
shrinking factor can be obtained directly from the optimal fidelity
as $\eta \left( N,M\right) =N\left( M+2\right) /M\left( N+2\right)
$. In this paper, we shall study the quantum broadcasting with input
as identical mixed qubits. We emphasize that the case considered in
this paper is the {\it universal quantum broadcasting which is in
the sense that the shrinking factor is independent from the input
state. Since we know that pure states set is a subset of that of
mixed states, and the optimal shrinking factor for mixed states
(also including pure state case) will be upper bounded by the
optimal shrinking factor for pure states}. Thus to show a
broadcasting of mixed states is optimal, we only need to show that
the shrinking factor is optimal. In this {\it universal} sense, we
mean that this quantum broadcasting procession can copy mixed qubits
equally well as it copies pure qubits.

Cirac \textit{et al} \cite{Cirac} introduced a decomposition of the
multi-qubit states of the form $\rho ^{\otimes N}$, where $\rho
=c_{1}|\uparrow \rangle \langle \uparrow |+|\downarrow \rangle
\langle \downarrow |$, $c_{1}+c_{0}=1$, and employed it to construct
the optimal single qubit purification procedure. The same
decomposition also was used to study the super-broadcasting of mixed states in Ref.%
\cite{Giacomo,Buscemi}. The states purification is also studied in
Ref.\cite{KW}. In this paper, we will provide a different pure state
decomposition for $\rho ^{\otimes N}$, and will present an optimal
{\it universal} quantum broadcasting for this case. For
completeness, we will also present the super-broadcasting for mixed
states which was studied in Refs.\cite{Cirac,Giacomo,Buscemi}. In
Ref.\cite{FLS}, the 2 to $M$ mixed states broadcasting was studied
in which no results about pure states decomposition of $\rho
^{\otimes N}$ is involved. We study in this paper the general $N$ to
$M$ mixed states universal broadcasting.

This paper is organized as follows. In Sec. II we introduce a new
decomposition method for $N$ identical qubits of mixed states. In
Sec. III we will show how to realize the optimal universal
broadcasting of mixed states from 3 to M copies by using the
orthonormal states introduced in Sec.
II. The quantum broadcasting transformation will be generalized to the case of $%
N\rightarrow M$ $\left( M\geq N\right) $ mixed-states in Sec IV. In
Sec. V, the super-broadcasting of mixed states will be presented.
Sec. VI contains several other slightly different quantum
broadcasting of mixed states. Sec. VII is a brief conclusion.

\section{Decomposition method for $N$ identical mixed states}

Before discussing the decomposition of multiqubit mixed states, we would
like to review briefly the multiqubit pure states. An arbitrary pure state
of two-dimensional system is described by the state vector
\begin{equation}
\left\vert \psi \right\rangle =a\left\vert \uparrow \right\rangle
+b\left\vert \downarrow \right\rangle ,  \label{1}
\end{equation}%
where $a$, $b$ are complex numbers and $\left\vert a\right\vert
^{2}+\left\vert b\right\vert ^{2}=1$. Therefore the $N$ identical pure
states can be written as%
\begin{equation}
\left\vert \psi \right\rangle ^{\otimes N}=\underset{m=0}{\overset{N}{\sum }}%
\sqrt{C_{N}^{m}}a^{N-m}b^{m}\left\vert \left( N-m\right) \uparrow
,m\downarrow \right\rangle ,  \label{2}
\end{equation}%
which is the superposition of all the N-qubit symmetric states, where $%
|(N-m)\uparrow ,m\downarrow \rangle $ is a symmetric state with
$N-m$ spins up and $m$ spins down, here we use the definition
$C_N^m\equiv \frac {N!}{m!(N-m)!}$. So, in the case of $N$ identical
\textit{pure} states as input states, we only need to consider
quantum cloning of symmetric states since we know $\left\vert \psi
\right\rangle ^{\otimes M}$ is still in the symmetric subspace. To
ensure each single qubit of the output state is the same, we can
restrict the output state of $M$-qubit in symmetric subspace.
Actually we can show that this cloning machine can still be optimal.
That is to say, the optimal universal quantum cloning of pure states
can only involve quantum states in symmetric subspace.

Now, let us consider the \textit{mixed} state input of $N$ identical qubits,
each qubit is an unknown state described by a density operator
\begin{equation}
\rho =c_{0}\left\vert \uparrow \right\rangle \left\langle \uparrow
\right\vert +c_{1}\left\vert \downarrow \right\rangle \left\langle
\downarrow \right\vert ,  \label{3}
\end{equation}%
where $c_{0}$ and $c_{1}$ are the probability of the distribution and $%
c_{1}+c_{0}=1$. Then the state of whole system, $\rho ^{\otimes N}$, can be
written as
\begin{eqnarray}
\rho ^{\otimes N} &=&\left( c_{0}\left\vert \uparrow \right\rangle
\left\langle \uparrow \right\vert +c_{1}\left\vert \downarrow \right\rangle
\left\langle \downarrow \right\vert \right) ^{\otimes N}  \notag \\
&=&\overset{N}{\underset{m=0}{\sum }}c_{0}^{N-m}c_{1}^{m}\overset{C_{N}^{m}}{%
\underset{j=1}{\sum }}\Pi _{j}\left[ \left( \left\vert \uparrow
\right\rangle \left\langle \uparrow \right\vert \right) ^{\otimes
N-m}\left( \left\vert \downarrow \right\rangle \left\langle
\downarrow \right\vert \right) ^{\otimes m}\right] ,  \label{4}
\end{eqnarray}%
where $\Pi _{j}$ denotes $j$-th permutation operator, $\Pi _{j}\in
S_{N}$, and the number of the permutation operators is $C_{N}^{m}$.
Inserting
identity $I=\overset{N}{\underset{m=0}{\sum }}\overset{C_{N}^{m}-1}{\underset%
{\alpha =0}{\sum }}\left\vert \left( N-m\right) \uparrow ,m\downarrow
\right\rangle _{\alpha \text{ }\alpha }\left\langle \left( N-m\right)
\uparrow ,m\downarrow \right\vert $ in Eq.(\ref{4}), we obtain%
\begin{eqnarray}
\rho ^{\otimes N} &=&\overset{N}{\underset{m=0}{\sum }}c_{0}^{N-m}c_{1}^{m}%
\overset{N}{\underset{m^{\prime }=0}{\sum }}\overset{C_{N}^{m^{\prime }}-1}{%
\underset{\alpha =0}{\sum }}\overset{N}{\underset{m^{\prime
\prime }=0}{\sum }}\overset{C_{N}^{m^{\prime \prime
}}-1}{\underset{\alpha ^{\prime }=0}{\sum }}\left\vert \left(
N-m^{\prime }\right) \uparrow ,m^{\prime }\downarrow
\right\rangle _{\alpha \text{ }\alpha ^{\prime }}\left\langle
\left( N-m^{\prime \prime }\right) \uparrow
,m^{\prime \prime }\downarrow \right\vert \notag \\
&&_{\alpha }\left\langle \left( N-m^{\prime }\right) \uparrow
,m^{\prime
}\downarrow \right\vert \overset{C_{N}^{m}}{\underset{j=1}{\sum }}\Pi _{j}%
\left[ \left( \left\vert \uparrow \right\rangle \left\langle
\uparrow \right\vert \right) ^{\otimes N-m}\left( \left\vert
\downarrow \right\rangle \left\langle \downarrow \right\vert
\right) ^{\otimes m}\right] \left\vert \left( N-m^{\prime \prime
}\right) \uparrow ,m^{\prime \prime }\downarrow \right\rangle
_{\alpha ^{\prime }}.  \label{5}
\end{eqnarray}

Here the orthonormal basis vector $\left\vert \left( N-m\right) \uparrow
,m\downarrow \right\rangle _{\alpha }$ is defined in terms of the states
with $(N-m)$ spins up and $m$ spins down, which is like a symmetric state
but with a phase in each term. The explicit expression is
\begin{eqnarray}
&&\left\vert \left( N-m\right) \uparrow ,m\downarrow \right\rangle _{\alpha }
\notag \\
&\equiv &\frac{1}{\sqrt{C_{N}^{m}}}\underset{j=1}{\overset{C_{N}^{m}}{\sum }}%
e^{2\pi i\alpha \left( j-1\right) /C_{N}^{m}}\Pi _{j}\left( \left\vert
\uparrow \right\rangle ^{\otimes N-m}\left\vert \downarrow \right\rangle
^{\otimes m}\right) .  \label{6}
\end{eqnarray}%
When $\alpha =0$ the quantum state $\left\vert \left( N-m\right) \uparrow
,m\downarrow \right\rangle _{0}\equiv \left\vert \left( N-m\right) \uparrow
,m\downarrow \right\rangle $ is symmetric state. Otherwise, the quantum
state $\left\vert \left( N-m\right) \uparrow ,m\downarrow \right\rangle
_{\alpha }$ is asymmetric because of different phases in each item. For
example, let $N=3$ and $m=1$, then $\alpha =0,1,2$. We have $\left\vert
2\uparrow ,\downarrow \right\rangle _{0}\equiv \left\vert 2\uparrow
,\downarrow \right\rangle =\frac{1}{\sqrt{3}}\left( \left\vert \uparrow
\uparrow \downarrow \right\rangle +\left\vert \uparrow \downarrow \uparrow
\right\rangle +\left\vert \downarrow \uparrow \uparrow \right\rangle \right)
$, $\left\vert 2\uparrow ,\downarrow \right\rangle _{1}=\frac{1}{\sqrt{3}}%
\left( \left\vert \uparrow \uparrow \downarrow \right\rangle +\omega
\left\vert \uparrow \downarrow \uparrow \right\rangle +\omega ^{2}\left\vert
\downarrow \uparrow \uparrow \right\rangle \right) $, $\left\vert 2\uparrow
,\downarrow \right\rangle _{2}=\frac{1}{\sqrt{3}}\left( \left\vert \uparrow
\uparrow \downarrow \right\rangle +\omega ^{2}\left\vert \uparrow \downarrow
\uparrow \right\rangle +\omega \left\vert \downarrow \uparrow \uparrow
\right\rangle \right) $, where $\omega =e^{2\pi i/3}$. It is obvious that $%
\left\vert 2\uparrow ,\downarrow \right\rangle _{1}$ and $\left\vert
2\uparrow ,\downarrow \right\rangle _{2}$ are not symmetric states under
permutation operator.

It is easy to check that Eq.(\ref{5}) is non-zero if and only if $%
m=m^{\prime }=m^{\prime \prime }$. In the subspace constructed of $N-m$
qubits in state $\left\vert \uparrow \right\rangle $ and $m$ qubits in state
$\left\vert \downarrow \right\rangle $, $\overset{C_{N}^{m}}{\underset{j=1}{%
\sum }}\Pi _{j}\left[ \left( \left\vert \uparrow \right\rangle \left\langle
\uparrow \right\vert \right) ^{\otimes N-m}\left( \left\vert \downarrow
\right\rangle \left\langle \downarrow \right\vert \right) ^{\otimes m}\right]
$ is an $C_{N}^{m}$-dimensional identity operator. Then we obtain%
\begin{eqnarray}
&&_{\alpha }\left\langle \left( N-m^{\prime }\right) \uparrow ,m^{\prime
}\downarrow \right\vert \overset{C_{N}^{m}}{\underset{j=1}{\sum }}\Pi _{j}%
\left[ \left( \left\vert \uparrow \right\rangle \left\langle
\uparrow \right\vert \right) ^{\otimes N-m}\left( \left\vert
\downarrow \right\rangle \left\langle \downarrow \right\vert
\right) ^{\otimes m}\right] \left\vert \left( N-m^{\prime \prime
}\right) \uparrow ,m^{\prime \prime }\downarrow \right\rangle
_{\alpha
^{\prime }}  \notag \\
&=&\delta _{m^{\prime }m}\delta _{mm^{\prime \prime }\text{ }\alpha
}\left\langle \left( N-m^{\prime }\right) \uparrow ,m^{\prime }\downarrow
\right\vert \left. \left( N-m^{\prime \prime }\right) \uparrow ,m^{\prime
\prime }\downarrow \right\rangle _{\alpha ^{\prime }}  \notag \\
&=&\delta _{m^{\prime }m}\delta _{mm^{\prime \prime }}\delta _{\alpha \alpha
^{\prime }}.  \label{7}
\end{eqnarray}%
Substituting Eq.(\ref{7}) into Eq.(\ref{5}), we obtain
\begin{eqnarray}
\rho ^{\otimes N} &=&\overset{N}{\underset{m=0}{\sum }}c_{0}^{N-m}c_{1}^{m}%
\overset{C_{N}^{m}-1}{\underset{\alpha =0}{\sum }}\left\vert
\left( N-m\right) \uparrow ,m\downarrow \right\rangle _{\alpha
\alpha }\left\langle \left( N-m\right) \uparrow ,m\downarrow
\right\vert .  \label{8}
\end{eqnarray}%
Eq.(\ref{8}) is the final decomposition result, in which the state of $N$
identical mixed qubit $\rho ^{\otimes N}$ is decomposed into the sum of pure
state density operators $\left\vert \left( N-m\right) \uparrow ,m\downarrow
\right\rangle _{\alpha \text{ }\alpha }\left\langle \left( N-m\right)
\uparrow ,m\downarrow \right\vert $. One important property of this state is
that each single-qubit reduced density operator is independent of the
subscript $\alpha $ and is the same to each other, which is similar as a
symmetric state.

In Ref.\cite{Cirac}, the authors introduced another decomposition method,
\begin{eqnarray}
\rho ^{\otimes N}=\overset{\frac{N}{2}}{\underset{j=\left\langle
\left\langle \frac{N}{2}\right\rangle \right\rangle }{\sum }}\overset{j}{%
\underset{m=-j}{\sum }}\overset{d_{j}}{\underset{\alpha =1}{\sum }}c_{0}^{%
\frac{N}{2}-m}c_{1}^{\frac{N}{2}+m}\left\vert jm\alpha \right\rangle
\left\langle jm\alpha \right\vert , \label{cirac}
\end{eqnarray}
where $\left\langle \left\langle \frac {N}{2}\right\rangle
\right\rangle $ is 0 for $N$ even, $1/2$ for $N$ odd, the detailed
definition of the notations in the above formula will be presented
in section VI. They also proposed a broadcasting procedure
for this state. The scheme is like the following: By unitary transformation $%
U_{j,\alpha }^{\dagger }\left\vert jm\alpha \right\rangle$, we obtain $%
\left\vert jm\right\rangle \otimes \widetilde{\left\vert \uparrow
,\downarrow \right\rangle }^{\otimes \frac{N}{2}-j}$, where
$\widetilde{\left\vert \uparrow ,\downarrow \right\rangle }
=(|\uparrow \downarrow \rangle -|\downarrow \uparrow \rangle )/\sqrt
{2}$ is the singlet state. Since the singlet state
$\widetilde{\left\vert \uparrow ,\downarrow \right\rangle }$ carries
no information about $\rho $, the last $N-2j$ qubits are discarded.
Then the rest $2j$ qubits in state $\left\vert jm\right\rangle $ are
cloned by using the known UQCM of pure states. As a whole, the input
states have different number of qubits and the number of output
qubits is the same. One may find that this quantum broadcasting may
not be universal. It is shown that the mixed states cannot be
broadcasted equally well as the pure states.

We will use a different scheme to study the broadcasting of mixed
states. For
example, if we use the pure state decomposition presented in Ref.\cite{Cirac}%
, we will broadcast the whole input state $|jm\alpha \rangle $ to
$M$ qubits. So the broadcasting procession does not depend on the
specified form of the input, that is, we do not consider which part
of the input should be cloned and which part should be discarded.
Thus the scalar relation for quantum broadcasting still holds,
\begin{equation}
\rho _{out}^{single}=\eta \rho +\frac{1-\eta }{2}I, ,  \label{9}
\end{equation}%
as mentioned previously, where $\eta $ is the shrinking factor which
is independent of input state $\rho $, and $I$ is the identity
operator.

As we know, the decomposition of $N$ identical mixed states $\rho
^{\otimes N}$ contains not only symmetric states but also asymmetric
states. For the symmetric states input, there have been optimal
universal cloning machine. However, the optimal universal quantum
broadcasting of asymmetric states has not been proposed. In the
following sections, by using our decomposition
introduced above, we will demonstrate how to broadcast the states $%
|(N-m)\uparrow ,m\downarrow \rangle _\alpha $ when $\alpha \not= 0$.
Thus we can realize the broadcasting of mixed states $\rho $ from
$N$ to $M\geq N$ copies. And we will also show that this
broadcasting is optimal and universal. To show explicit how we
broadcast the mixed qubits, we present in detail the broadcasting
transformations for the 3 to $M$ case.

\section{$3\longrightarrow M$ qubits quantum broadcasting of mixed states}

The result of $2$ to $M$ optimal universal quantum broadcasting
of mixed states have been studied by Fan \textit{et al} in Ref.\cite%
{FLS}. In this section, we demonstrate in detail how to realize the
optimal universal broadcasting from $3$ to $M$ copies of mixed
states.

As we shown, an arbitrary mixed state of two-dimensional system takes the
form
\begin{equation}
\rho =c_{0}\left\vert \uparrow \right\rangle \left\langle \uparrow
\right\vert +c_{1}\left\vert \downarrow \right\rangle \left\langle
\downarrow \right\vert ,  \label{10}
\end{equation}
where $c_{0}+c_{1}=1$. According to Eq.(\ref{8}), the three identical mixed
states can be decomposed as
\begin{align}
\rho ^{\otimes 3}& =c_{0}^{3}\left\vert 3\uparrow \right\rangle \left\langle
3\uparrow \right\vert +c_{0}^{2}c_{1}\left\vert 2\uparrow ,\downarrow
\right\rangle \left\langle 2\uparrow ,\downarrow \right\vert  \notag \\
& +c_{0}c_{1}^{2}\left\vert \uparrow ,2\downarrow \right\rangle \left\langle
\uparrow ,2\downarrow \right\vert +c_{1}^{3}\left\vert 3\downarrow
\right\rangle \left\langle 3\downarrow \right\vert  \notag \\
& +c_{0}^{2}c_{1}\left\vert 2\uparrow ,\downarrow \right\rangle _{1\text{ }%
1}\left\langle 2\uparrow ,\downarrow \right\vert +c_{0}c_{1}^{2}\left\vert
\uparrow ,2\downarrow \right\rangle _{1\text{ }1}\left\langle \uparrow
,2\downarrow \right\vert  \notag \\
& +c_{0}^{2}c_{1}\left\vert 2\uparrow ,\downarrow \right\rangle _{2\text{ }%
2}\left\langle 2\uparrow ,\downarrow \right\vert
+c_{0}c_{1}^{2}\left\vert \uparrow ,2\downarrow \right\rangle
_{2\text{ }2}\left\langle \uparrow ,2\downarrow \right\vert .
\label{11}
\end{align}%
It is well known that the quantum cloning transformations of
symmetric states are described as \cite{Fan},
\begin{subequations}
\begin{equation}
U_{3M}\left\vert 3\uparrow \right\rangle \otimes R=\overset{M-3}{\underset{%
k=0}{\sum }}\beta _{0k}\left\vert \left( M-k\right) \uparrow ,k\downarrow
\right\rangle \otimes R_{\left\vert \left( M-3-k\right) \uparrow
,k\downarrow \right\rangle },  \label{12a}
\end{equation}%
\begin{eqnarray}
U_{3M}\left\vert 2\uparrow ,\downarrow \right\rangle \otimes R &=&\overset{%
M-3}{\underset{k=0}{\sum }}\beta _{1k}\left\vert \left(
M-1-k\right) \uparrow ,\left( k+1\right) \downarrow
\right\rangle  \otimes R_{\left\vert \left( M-3-k\right) \uparrow
,k\downarrow \right\rangle },  \label{12b}
\end{eqnarray}%
\begin{eqnarray}
U_{3M}\left\vert \uparrow ,2\downarrow \right\rangle \otimes R &=&\overset{%
M-3}{\underset{k=0}{\sum }}\beta _{2k}\left\vert \left(
M-2-k\right) \uparrow ,\left( k+2\right) \downarrow
\right\rangle  \otimes R_{\left\vert \left( M-3-k\right) \uparrow
,k\downarrow \right\rangle },  \label{12c}
\end{eqnarray}%
\begin{eqnarray}
U_{3M}\left\vert 3\downarrow \right\rangle \otimes R &=&\overset{M-3}{%
\underset{k=0}{\sum }}\beta _{3k}\left\vert \left( M-3-k\right)
\uparrow ,\left( k+3\right) \downarrow \right\rangle \otimes
R_{\left\vert \left( M-3-k\right) \uparrow ,k\downarrow
\right\rangle }.  \label{12d}
\end{eqnarray}%
where $R$ denotes the blank state and the ancillary state, $R_{|\psi
\rangle } $ denotes the ancillary state which should be traced out
to obtain the output of the quantum broadcasting, in principle, the
ancillary state is the same as the state $|\psi \rangle $. We write
it as the subscript to distinguish the ancillary state from the
copied state. By the cloning transformations
Eq.(\ref{12a})-Eq.(\ref{12d}), the output qubits are the same
because the output states are $M$-qubit symmetric states.

As for the states with $\alpha \not= 0$, we propose the broadcasting
transformations as below,%
\begin{eqnarray}
U_{3M}\left\vert 2\uparrow ,\downarrow \right\rangle _{1}\otimes R &=&%
\overset{M-3}{\underset{k=0}{\sum }}\beta _{1k}\left\vert \left(
M-1-k\right) \uparrow ,\left( k+1\right) \downarrow \right\rangle
_{1} \otimes R_{\left\vert \left( M-3-k\right) \uparrow
,k\downarrow \right\rangle _{1}},  \label{12e}
\end{eqnarray}%
\begin{eqnarray}
U_{3M}\left\vert \uparrow ,2\downarrow \right\rangle _{1}\otimes R &=&%
\overset{M-3}{\underset{k=0}{\sum }}\beta _{2k}\left\vert \left(
M-2-k\right) \uparrow ,\left( k+2\right) \downarrow \right\rangle
_{1} \otimes R_{\left\vert \left( M-3-k\right) \uparrow
,k\downarrow \right\rangle _{1}},  \label{12f}
\end{eqnarray}%
\begin{eqnarray}
U_{3M}\left\vert 2\uparrow ,\downarrow \right\rangle _{2}\otimes R &=&%
\overset{M-3}{\underset{k=0}{\sum }}\beta _{1k}\left\vert \left(
M-1-k\right) \uparrow ,\left( k+1\right) \downarrow \right\rangle
_{2} \otimes R_{\left\vert \left( M-3-k\right) \uparrow
,k\downarrow \right\rangle _{2}},  \label{12g}
\end{eqnarray}%
\begin{eqnarray}
U_{3M}\left\vert \uparrow ,2\downarrow \right\rangle _{2}\otimes R &=&%
\overset{M-3}{\underset{k=0}{\sum }}\beta _{2k}\left\vert \left(
M-2-k\right) \uparrow ,\left( k+2\right) \downarrow \right\rangle
_{2} \otimes R_{\left\vert \left( M-3-k\right) \uparrow
,k\downarrow \right\rangle _{2}},  \label{12h}
\end{eqnarray}%
where, as we mentioned,  $R_{\left\vert \left( M-3-k\right) \uparrow
,k\downarrow \right\rangle _{\alpha }}$ $\left( \alpha =0,1,2\right)
$ are ancillary states and
orthogonal to each other. A simple realization of them is that $%
R_{\left\vert \left( M-3-k\right) \uparrow ,k\downarrow \right\rangle
_{\alpha }}= \left\vert \left( M-3-k\right) \uparrow ,k\downarrow
\right\rangle _{\alpha }$. $\beta _{mk}$ is the proposed amplitude for each
item, and has the following form,
\end{subequations}
\begin{eqnarray}
\beta _{mk} &=&\sqrt{\frac{\left( M-3\right) !\left( 3+1\right)
!}{\left( M+1\right) !}}\sqrt{\frac{\left( M-m-k\right) !}{\left(
3-m\right) !\left( M-3-k\right) !}}  \sqrt{\frac{\left(
m+k\right) !}{m!k!}},\text{ } \label{13}
\end{eqnarray}%
where $m=0,1,2,3$. The reason why we propose this kind of
broadcasting transformation is that we would like to let input
states $|(N-m)\uparrow ,m\downarrow \rangle _{\alpha }$ have the
same output reduced density operators for different $\alpha $. As we
already know the result for $\alpha =0$ in Ref.\cite{Fan}, naturally
we propose the same $\beta _{mk}$ for different $\alpha $. We note
that the three-qubit asymmetric states are transformed into
$M$-qubit asymmetric states, which have the same subscripts after
the broadcasting transformation. However, they have different phases
by definition in each superposition item. It is easy to check that
the relations in Eq.(\ref{12a})-(\ref{12h}) satisfy the unitary
condition. By tracing out ancillary states $R_{|\psi \rangle }$
denoted as $Rs$, we obtain the output state of $M$ qubits as below.
\begin{eqnarray}
\rho _{out}&=&Tr_{Rs}\left[ U_{3M}\left( \rho ^{\otimes 3}\otimes
R\right)
U_{3M}^{\dagger }\right] \notag \\
&=&c_{0}^{3}\overset{M-3}{\underset{k=0}{\sum }}\beta
_{0k}^{2}\left\vert \left( M-k\right) \uparrow ,k\downarrow
\right\rangle \left\langle \left(
M-k\right) \uparrow ,k\downarrow \right\vert \notag \\
&&+c_{0}^{2}c_{1}\overset{M-3}{\underset{k=0}{\sum }}\beta
_{1k}^{2}\left\vert \left( M-1-k\right) \uparrow ,\left(
k+1\right) \downarrow \right\rangle \left\langle \left(
M-1-k\right) \uparrow ,\left( k+1\right) \downarrow
\right\vert \notag \\
&&+c_{0}c_{1}^{2}\overset{M-3}{\underset{k=0}{\sum }}\beta
_{2k}^{2}\left\vert \left( M-2-k\right) \uparrow ,\left(
k+2\right) \downarrow \right\rangle \left\langle \left(
M-2-k\right) \uparrow ,\left( k+2\right) \downarrow
\right\vert \notag \\
&&+c_{1}^{3}\overset{M-3}{\underset{k=0}{\sum }}\beta
_{3k}\left\vert \left( M-3-k\right) \uparrow ,\left( k+3\right)
\downarrow \right\rangle \left\langle \left( M-3-k\right)
\uparrow ,\left( k+3\right) \downarrow
\right\vert \notag \\
&&+c_{0}^{2}c_{1}\overset{M-3}{\underset{k=0}{\sum }}\beta
_{1k}^{2}\left\vert \left( M-1-k\right) \uparrow ,\left(
k+1\right) \downarrow \right\rangle _{11}\left\langle \left(
M-1-k\right) \uparrow ,\left( k+1\right)
\downarrow \right\vert \notag \\
&& +c_{0}c_{1}^{2}\overset{M-3}{\underset{k=0}{\sum }}\beta
_{2k}^{2}\left\vert \left( M-2-k\right) \uparrow ,\left(
k+2\right) \downarrow \right\rangle _{11}\left\langle \left(
M-2-k\right) \uparrow ,\left( k+2\right) \downarrow \right\vert
\notag \\
&&+c_{0}^{2}c_{1}\overset{M-3}{\underset{k=0}{\sum
}}\beta _{1k}^{2}\left\vert \left( M-1-k\right) \uparrow ,\left(
k+1\right) \downarrow \right\rangle _{22}\left\langle \left(
M-1-k\right) \uparrow ,\left( k+1\right)
\downarrow \right\vert  \notag \\
&&+c_{0}c_{1}^{2}\overset{M-3}{\underset{k=0}{\sum }}\beta
_{2k}^{2}\left\vert \left( M-2-k\right) \uparrow ,\left(
k+2\right) \downarrow \right\rangle _{22}\left\langle \left(
M-2-k\right) \uparrow ,\left( k+2\right) \downarrow \right\vert
.  \label{14}
\end{eqnarray}

To evaluate the quality of output qubits, we should compare the
single-qubit reduced density operator of output state with the input
state $\rho $. As we know that in our decomposition, for a given
multi-qubit state, symmetric or asymmetric (in a specified form as
we defined), the single-qubit reduced density operators are the same
to each other and are unrelated with symmetry. Then we have the
following relations,
\begin{eqnarray}
&&Tr_{M-1}\left[ \left\vert \left( M-m-k\right) \uparrow ,\left(
m+k\right) \downarrow \right\rangle \left\langle \left(
M-m-k\right) \uparrow ,\left( m+k\right)
\downarrow \right\vert \right]  \notag \\
&=&\frac{C_{M-1}^{m+k}}{C_{M}^{m+k}}\left\vert \uparrow \right\rangle
\left\langle \uparrow \right\vert +\frac{C_{M-1}^{m+k-1}}{C_{M}^{m+k}}%
\left\vert \downarrow \right\rangle \left\langle \downarrow \right\vert
\notag \\
&=&\frac{M-m-k}{M}\left\vert \uparrow \right\rangle \left\langle
\uparrow \right\vert +\frac{m+k}{M}\left\vert \downarrow
\right\rangle \left\langle \downarrow \right\vert ,  ~~~~~~\left(
m=0,1,2,3\right)  \label{15}
\end{eqnarray}%
and%
\begin{eqnarray}
&&Tr_{M-1}\left[ \left\vert \left( M-m-k\right) \uparrow ,\left(
m+k\right) \downarrow \right\rangle \left\langle \left(
M-m-k\right) \uparrow ,\left( m+k\right)
\downarrow \right\vert \right]  \notag \\
&=&Tr_{M-1}\left[ \left\vert \left( M-m-k\right) \uparrow ,\left(
m+k\right) \downarrow \right\rangle _{\alpha \alpha }\left\langle
\left( M-m-k\right) \uparrow ,\left( m+k\right) \downarrow
\right\vert \right] ,  \label{16}
\end{eqnarray}
where $m=1,2$ and $\alpha =1,2$. Using Eq.(\ref{15}) and Eq.(\ref{16}), we
obtain the scalar relation for single-qubit reduced density operator,%
\begin{equation}
\rho _{out}^{single}=Tr_{M-1}\left( \rho _{out}\right) =\frac{3\left(
M+2\right) }{5M}\rho +\frac{M-3}{5M}I.  \label{17}
\end{equation}%
It is obvious that the shrinking factor is independent of input state and
reaches the upper bound $\frac{3\left( M+2\right) }{5M}$. So, we find the $%
3\rightarrow M$ quantum broadcasting of mixed states which is
universal, symmetric and optimal.

\section{Generalization: $N\longrightarrow M$ qubits quantum broadcasting of
mixed states}

Here, we consider the general broadcasting case, that is, creating
$M$ qubits from $N$ identical mixed states ($M\geq N$). Let the
input state be $N$ identical qubits, each in an unknown state
described by a density operator $\rho $.
According to Eq.(\ref{8}), the decomposition of $\rho ^{\otimes N}$ reads%
\begin{eqnarray}
\rho ^{\otimes N} =\overset{N}{\underset{m=0}{\sum }}c_{0}^{N-m}c_{1}^{m}%
\overset{C_{N}^{m}-1}{\underset{\alpha =0}{\sum }}\left\vert
\left( N-m\right) \uparrow ,m\downarrow \right\rangle _{\alpha
\alpha }\left\langle \left( N-m\right) \uparrow ,m\downarrow
\right\vert .  \label{18}
\end{eqnarray}

We propose the general quantum broadcasting transformation from $N$
to $M$ qubits as follows,
\begin{eqnarray}
&&U_{NM}\left[ \left\vert \left( N-m\right) \uparrow ,m\downarrow
\right\rangle _{\alpha }\otimes R\right]  \notag \\
&=&\overset{M-N}{\underset{k=0}{\sum }}\beta _{mk}\left\vert
\left( M-m-k\right) \uparrow ,\left( m+k\right) \downarrow
\right\rangle _{\alpha } \otimes R_{\left\vert \left(
M-N-k\right) \uparrow ,k\downarrow \right\rangle _{\alpha }},
\label{19}
\end{eqnarray}%
where%
\begin{eqnarray}
\beta _{mk} =\sqrt{\frac{\left( M-N\right) !\left( N+1\right)
!}{\left( M+1\right) !}}\sqrt{\frac{\left( M-m-k\right) !}{\left(
N-m\right) !\left( M-N-k\right) !}} \sqrt{\frac{\left( m+k\right)
!}{m!k!}}.  \label{20}
\end{eqnarray}%
The unitary operator $U_{NM}$ denotes the map from $N$ identical qubits to $%
M $ qubits. $\beta _{mk}$ are amplitudes for each item. In
Eq.(\ref{19}), we
still let the input state and output state have the same subscript. From Eq.(%
\ref{6}), one can find their specific expressions.

Our aim is to show that the proposed broadcasting transformation is
universal and optimal. So we should show that the scalar relation is
satisfied and in it the shrinking factor does not depend on the
input, that means this broadcasting is universal. At the same time,
if the shrinking factor saturates its upper bound, we mean that this
broadcasting procession is optimal.

By tracing out $R$, we obtain the single qubit output density operator%
\begin{eqnarray}
\rho _{out}&=&Tr_{R}\left[ U_{NM}\left( \rho ^{\otimes N}\otimes
R\right)
U_{NM}^{\dagger }\right]  \notag \\
&=&\overset{N}{\underset{m=0}{\sum }}c_{0}^{N-m}c_{1}^{m}\overset{C_{N}^{m}-1%
}{\underset{\alpha =0}{\sum }}Tr_{R}\left\{ U_{NM}\left[
\left\vert \left( N-m\right) \uparrow ,m\downarrow \right\rangle
_{\alpha \alpha }\left\langle \left( N-m\right) \uparrow
,m\downarrow \right\vert \otimes R\right] U_{NM}^{\dagger
}\right\}  \notag
\\
&=&\overset{N}{\underset{m=0}{\sum }}c_{0}^{N-m}c_{1}^{m}\overset{C_{N}^{m}-1%
}{\underset{\alpha =0}{\sum }}\underset{k=0}{\overset{M-N}{\sum
}}\beta _{mk}^{2}\left\vert \left( M-m-k\right) \uparrow ,\left(
m+k\right) \downarrow \right\rangle _{\alpha \text{ }\alpha
}\left\langle \left( M-m-k\right) \uparrow ,\left( m+k\right)
\downarrow \right\vert .  \label{21}
\end{eqnarray}%
Here the linearity superposition of quantum states is used. Then the output
single-qubit reduced density operator reads%
\begin{eqnarray}
\rho _{out}^{single}&=&Tr_{M-1}\left( \rho _{out}\right)  \notag \\
&=&\overset{N}{\underset{m=0}{\sum }}c_{0}^{N-m}c_{1}^{m}\overset{C_{N}^{m}-1%
}{\underset{\alpha =0}{\sum }}\underset{k=0}{\overset{M-N}{\sum }}\beta
_{mk}^{2}Tr_{M-1}\left[ \left\vert \left( M-m-k\right) \uparrow ,
\left( m+k\right) \downarrow \right\rangle _{\alpha \text{ }%
\alpha }\left\langle \left( M-m-k\right) \uparrow ,\left( m+k\right)
\downarrow \right\vert \right]  \notag \\
&=&\overset{N}{\underset{m=0}{\sum }}c_{0}^{N-m}c_{1}^{m}\overset{C_{N}^{m}-1%
}{\underset{\alpha =0}{\sum }}\underset{k=0}{\overset{M-N}{\sum
}}\beta _{mk}^{2}\left( \frac{M-m-k}{M}\left\vert \uparrow
\right\rangle \left\langle \uparrow \right\vert
+\frac{m+k}{M}\left\vert \downarrow \right\rangle \left\langle
\downarrow \right\vert \right)  \notag \\
&=&\frac{N\left( M+2\right) }{M\left( N+2\right) }\rho
+\frac{M-N}{M\left( N+2\right) }I.  \label{22}
\end{eqnarray}%
To calculate Eq.(\ref{22}), the following relations are used.%
\begin{eqnarray}
&& Tr_{M-1}\left[ \left\vert \left( M-m-k\right) \uparrow ,\left(
m+k\right) \downarrow \right\rangle _{\alpha \alpha }\left\langle
\left( M-m-k\right) \uparrow ,\left(
m+k\right) \downarrow \right\vert \right]  \notag \\
&=&\frac{C_{M-1}^{m+k}}{C_{M}^{m+k}}\left\vert \uparrow
\right\rangle
\left\langle \uparrow \right\vert +\frac{C_{M-1}^{m+k-1}}{C_{M}^{m+k}}%
\left\vert \downarrow \right\rangle \left\langle \downarrow \right\vert
\notag \\
&=&\frac{M-m-k}{M}\left\vert \uparrow \right\rangle \left\langle
\uparrow \right\vert +\frac{m+k}{M}\left\vert \downarrow
\right\rangle \left\langle \downarrow \right\vert .  \label{23}
\end{eqnarray}

Obviously scalar relation between single qubit input $\rho $ and
single-qubit output reduced density operator is satisfied, $\rho
_{out}^{single}=\frac{N\left( M+2\right) }{M\left( N+2\right) }\rho +\frac{%
M-N}{M\left( N+2\right) }I$, and we find the shrinking factor
$\frac{N\left( M+2\right) }{M\left( N+2\right) }$ achieves the well
known upper bound. Thus our proposed quantum broadcasting is
optimal. Since the upper bound is saturated, we know this
broadcasting procession can copy mixed state equally well as the
pure state. In short we demonstrated an optimal quantum broadcasting
which can transform $N$ identical mixed states into $M$-qubit states
$\left( M\geqslant N\right) $.

\section{Other optimal universal quantum broadcasting procedures for mixed states}

Besides the above mentioned quantum broadcasting, there are several
other quantum broadcasting procedures which are still optimal.
In Eq.(\ref{19}), the $N$-qubit symmetric states are transformed to $%
M$-qubit symmetric states, while states $|(N-m)\uparrow ,m\downarrow
\rangle _{\alpha }$ are transformed to $M$-qubit states with similar
form. In the process of this quantum broadcasting, the input state
and output state have the same subscript $\alpha $. However, we
found that when $\left\vert \left( N-m\right) \uparrow ,m\downarrow
\right\rangle _{\alpha }$ is transformed
into state $\frac{1}{\sqrt{C_{M}^{m+k}}}\overset{C_{M}^{m+k}-1}{\underset{%
\alpha =0}{\sum }}\left\vert \left( M-m-k\right) \uparrow ,\left(
m+k\right) \downarrow \right\rangle _{\alpha }$, it is also an
optimal universal broadcasting transformation. The transformation is
presented as follows,
\begin{eqnarray}
&&U_{NM}^{\prime }\left\vert \left( N-m\right) \uparrow ,m\downarrow
\right\rangle _{\alpha }\otimes R  \notag \\
&=&\overset{M-N}{\underset{k=0}{\sum }}\beta _{mk}\frac{1}{\sqrt{C_{M}^{m+k}}%
}\overset{C_{M}^{m+k}-1}{\underset{\alpha ^{\prime }=0}{\sum
}}\left\vert \left( M-m-k\right) \uparrow ,\left( m+k\right)
\downarrow \right\rangle _{\alpha ^{\prime }}\otimes
R_{\left\vert \left( M-N-k\right) \uparrow ,k\downarrow
\right\rangle _{\alpha ^{\prime }}}.  \label{24}
\end{eqnarray}
where terms in the superposition of the output state are changed to another
form compared with the results in Section IV.

If we use the decomposition method in Ref.\cite{Cirac}, it is still
possible to realize other two quantum broadcastings. The first one
takes the form
\begin{eqnarray}
U_{NM}^{\prime \prime }\left\vert jm\alpha \right\rangle \otimes
R =\overset{M-N}{\underset{k=0}{\sum }}\Upsilon _{mk}\left\vert \left( M-%
\frac{N}{2}-m-k\right) \uparrow ,\left( \frac{N}{2}+m+k\right)
\downarrow \right\rangle _{\alpha }\otimes R_{\left\vert \left(
M-N-k\right) \uparrow ,k\downarrow \right\rangle _{\alpha }}.
\label{25}
\end{eqnarray}
The second one takes the form
\begin{eqnarray}
&&U_{NM}^{\prime \prime \prime }\left\vert jm\alpha \right\rangle
\otimes R
\notag \\
&=&\overset{M-N}{\underset{k=0}{\sum }}\Upsilon _{mk}\frac{1}{\sqrt{%
C_{M}^{m+k}}}\overset{C_{M}^{m+k}-1}{\underset{\alpha ^{\prime }=0}{\sum }}%
\left\vert \left( M-\frac{N}{2}-m-k\right) \uparrow ,\left(
\frac{N}{2}+m+k\right) \downarrow \right\rangle _{\alpha ^{\prime
}}\otimes R_{\left\vert \left( M-N-k\right) \uparrow ,k\downarrow
\right\rangle _{\alpha ^{\prime }}}, \label{26}
\end{eqnarray}%
where the normalization factor%
\begin{eqnarray}
\Upsilon _{mk} =\sqrt{\frac{\left( M-N\right) !\left( N+1\right)
!}{\left(
M+1\right) !}}\sqrt{\frac{\left( M-\frac{N}{2}-m-k\right) !}{\left( \frac{N}{%
2}-m\right) !\left( M-N-k\right) !}}\sqrt{\frac{\left(
\frac{N}{2}+m+k\right) !}{\left( \frac{N}{2}+m\right) !k!}}.
\label{27}
\end{eqnarray}

It is straightforward to check that the above quantum broadcasting
procedures are universal, symmetric and optimal. That is to say, for
$N$ copies of given mixed state $\rho $ there exist not only one
optimal universal quantum broadcasting procedures.

\section{$N\longrightarrow M$ copying by superbroadcasting scheme}
Cirac {\it et al} proposed an identical mixed states purification
scheme followed by a pure state cloning machine \cite{Cirac}, and
this is later related with the mixed states superbroadcasting
\cite{Giacomo,Buscemi}. Here, we review those results and compare
them with the results in this paper.  Let the input state be $N$
identical qubits, each in an unknown state described by a density
operator $\rho =c_{0}\left\vert \uparrow \right\rangle
\left\langle \uparrow \right\vert +c_{1}\left\vert \downarrow
\right\rangle \left\langle \downarrow \right\vert $, where
$c_{0}+c_{1}=1$. According to Eq.(\ref{cirac}), the
decomposition of $\rho ^{\otimes N}$ reads%
\begin{equation}
\rho ^{\otimes N}=\overset{\frac{N}{2}}{\underset{j=\left\langle
\left\langle \frac{N}{2}\right\rangle \right\rangle }{\sum }}\overset{j}{%
\underset{m=-j}{\sum }}\overset{d_{j}}{\underset{\alpha =1}{\sum }}c_{0}^{%
\frac{N}{2}-m}c_{1}^{\frac{N}{2}+m}\left\vert jm\alpha
\right\rangle \left\langle jm\alpha \right\vert ,
\end{equation}%
where%
\begin{equation}
d_{j}=C_{N}^{\frac{N}{2}-j}-C_{N}^{\frac{N}{2}-j-1},
\end{equation}%
\begin{equation}
\left\vert jm\alpha \right\rangle =U_{j,\alpha }\left\vert
jm1\right\rangle =U_{j,\alpha }\left[ \left\vert \left(
j-m\right) \uparrow ,\left( j+m\right) \downarrow \right\rangle
\otimes \widetilde{\left\vert \uparrow ,\downarrow \right\rangle
}^{\otimes \frac{N}{2}-j}\right] .
\end{equation}
Let $\alpha =0$ for $j=\frac{N}{2}$, and $\alpha =1,2,3,...,d_{j}$
when $j\neq \frac{N}{2}$.

The detailed super-broadcasting scheme is shown as follows. Firstly,
we measure the input $N$ qubits $\rho ^{\otimes N}$, if getting
state $\left\vert jm\alpha \right\rangle $, we take an unitary
operation $U_{j,\alpha }$ and change it into state $\left\vert
jm\right\rangle \otimes \widetilde{\left\vert \uparrow ,\downarrow
\right\rangle }^{\otimes \frac{N}{2}-j}=\left\vert \left( j-m\right)
\uparrow ,\left( j+m\right) \downarrow \right\rangle \otimes
\widetilde{\left\vert \uparrow ,\downarrow \right\rangle }^{\otimes
\frac{N}{2}-j}$. Then we discard the last $N-2j$ qubits, because
these qubits are singlets without any information of $\rho $.
Secondly, we use the optimal universal cloning transformations of
pure states to clone $2j$ qubits in symmetric state to $M$ output
qubits.
\begin{equation}
U_{2jM}\left\vert jm\right\rangle \otimes R=\overset{M-2j}{\underset{k=0}{%
\sum }}\beta _{mk}\left\vert \left( M-j-m-k\right) \uparrow
,\left( j+m+k\right) \downarrow \right\rangle \otimes R_{\left(
M-2j-k\right) \uparrow ,k\downarrow },
\end{equation}%
where%
\begin{equation}
\beta _{mk}=\sqrt{\frac{\left( M-2j\right) !\left( 2j+1\right)
!}{\left( M+1\right) !}}\sqrt{\frac{\left( M-j-m-k\right)
!}{\left( j-m\right) !\left( M-2j-k\right) !}\frac{\left(
j+m+k\right) !}{\left( j+m\right) !k!}}.
\end{equation}
The unitary operator $U_{2jM}$ denotes the map from $2j$ qubits to
$M$ qubits. By tracing out ancillary states $Rs$, we obtain the output density operator%
\begin{align}
\rho _{out}& =Tr_{Rs}\left[
\overset{\frac{N}{2}}{\underset{j=\left\langle
\left\langle \frac{N}{2}\right\rangle \right\rangle }{\sum }}\overset{j}{%
\underset{m=-j}{\sum }}c_{0}^{\frac{N}{2}-m}c_{1}^{\frac{N}{2}%
+m}d_{j}U_{2jM}\left( \left\vert jm\right\rangle \left\langle
jm\right\vert
\otimes R\right) U_{2jM}^{\dagger }\right]  \notag \\
& =\overset{\frac{N}{2}}{\underset{j=\left\langle \left\langle \frac{N}{2}%
\right\rangle \right\rangle }{\sum }}\overset{j}{\underset{m=-j}{\sum }}%
c_{0}^{\frac{N}{2}-m}c_{1}^{\frac{N}{2}+m}d_{j}\overset{M-2j}{\underset{k=0}{%
\sum }}\beta _{mk}^{2}  \notag \\
& \times \left\vert \left( M-j-m-k\right) \uparrow ,\left(
j+m+k\right) \downarrow \right\rangle \left\langle \left(
M-j-m-k\right) \uparrow ,\left( j+m+k\right) \downarrow \right\vert
.
\end{align}%
By using%
\begin{eqnarray*}
&&Tr_{M-1}\left[ \left\vert \left( M-j-m-k\right) \uparrow ,\left(
j+m+k\right) \downarrow \right\rangle j\left\langle \left(
M-j-m-k\right)
\uparrow ,\left( j+m+k\right) \downarrow \right\vert \right] \\
&=&\frac{C_{M-1}^{j+m+k}}{C_{M}^{j+m+k}}\left\vert \uparrow
\right\rangle
\left\langle \uparrow \right\vert +\frac{C_{M-1}^{j+m+k-1}}{C_{M}^{j+m+k}}%
\left\vert \downarrow \right\rangle \left\langle \downarrow
\right\vert \nonumber \\
&=& \frac{M-j-m-k}{M}\left\vert \uparrow \right\rangle \left\langle
\uparrow \right\vert +\frac{j+m+k}{M}\left\vert \downarrow
\right\rangle \left\langle \downarrow \right\vert ,
\end{eqnarray*}%
the output single-qubit reduced density operator reads%
\begin{eqnarray}
\rho _{single}^{out} &=&Tr_{M-1}\left( \rho _{out}\right) \nonumber
\\
&=&\overset{\frac{N}{2%
}}{\underset{j=\left\langle \left\langle \frac{N}{2}\right\rangle
\right\rangle }{\sum }}\overset{j}{\underset{m=-j}{\sum }}c_{0}^{\frac{N}{2}%
-m}c_{1}^{\frac{N}{2}+m}d_{j}\overset{M-2j}{\underset{k=0}{\sum
}}\beta _{mk}^{2}\left( \frac{M-j-m-k}{M}\left\vert \uparrow
\right\rangle \left\langle \uparrow \right\vert
+\frac{j+m+k}{M}\left\vert \downarrow
\right\rangle \left\langle \downarrow \right\vert \right)  \nonumber \\
&=&c_{0}^{\prime \prime }\left\vert \uparrow \right\rangle
\left\langle \uparrow \right\vert +c_{1}^{\prime \prime }\left\vert
\downarrow \right\rangle \left\langle \downarrow \right\vert
\nonumber \\
&=&\frac{1}{2}\left( I+r^{\prime \prime
}\overrightarrow{n}\overrightarrow{\sigma }\right) \nonumber \\
&=&r^{\prime \prime }\left\vert \uparrow \right\rangle \left\langle
\uparrow \right\vert +\frac{1-r^{\prime \prime }}{2}I,
\end{eqnarray}%
where
\begin{equation}
c_{0}^{\prime \prime
}=\overset{\frac{N}{2}}{\underset{j=\left\langle
\left\langle \frac{N}{2}\right\rangle \right\rangle }{\sum }}\overset{j}{%
\underset{m=-j}{\sum }}c_{0}^{\frac{N}{2}-m}c_{1}^{\frac{N}{2}+m}d_{j}%
\overset{M-2j}{\underset{k=0}{\sum }}\beta
_{mk}^{2}\frac{M-j-m-k}{M},
\end{equation}%
\begin{equation}
c_{1}^{\prime \prime
}=\overset{\frac{N}{2}}{\underset{j=\left\langle
\left\langle \frac{N}{2}\right\rangle \right\rangle }{\sum }}\overset{j}{%
\underset{m=-j}{\sum }}c_{0}^{\frac{N}{2}-m}c_{1}^{\frac{N}{2}+m}d_{j}%
\overset{M-2j}{\underset{k=0}{\sum }}\beta
_{mk}^{2}\frac{j+m+k}{M},
\end{equation}%
\begin{eqnarray}
r^{\prime \prime } &=&c_{0}^{\prime \prime }-c_{1}^{\prime \prime }
\nonumber \\
&=&\overset{%
\frac{N}{2}}{\underset{j=\left\langle \left\langle
\frac{N}{2}\right\rangle
\right\rangle }{\sum }}\overset{j}{\underset{m=-j}{\sum }}c_{0}^{\frac{N}{2}%
-m}c_{1}^{\frac{N}{2}+m}d_{j}\overset{M-2j}{\underset{k=0}{\sum
}}\beta
_{mk}^{2}\frac{M-2\left( j+m+k\right) }{M}  \notag \\
&=&\overset{\frac{N}{2}}{\underset{j=\left\langle \left\langle \frac{N}{2}%
\right\rangle \right\rangle }{\sum }}\overset{j}{\underset{m=-j}{\sum }}%
c_{0}^{\frac{N}{2}-m}c_{1}^{\frac{N}{2}+m}d_{j}\left[
\frac{-\left(
M+2\right) m}{M\left( j+1\right) }\right]  \notag \\
&=&-\frac{M+2}{M}\overset{\frac{N}{2}}{\underset{j=\left\langle
\left\langle \frac{N}{2}\right\rangle \right\rangle }{\sum
}}\frac{d_{j}}{\left(
j+1\right) }\overset{j}{\underset{m=-j}{\sum }}\left( \frac{1+r}{2}\right) ^{%
\frac{N}{2}-m}\left( \frac{1-r}{2}\right) ^{\frac{N}{2}+m}m.
\end{eqnarray}

Obviously the output single-qubit reduced density operator takes the form%
\begin{equation}
\rho _{single}^{out}=\frac{r^{\prime \prime }}{r}\rho
+\frac{1-r^{\prime \prime }/r}{2}I,
\end{equation}%
where the shrinking factor is $\eta =\frac{r^{\prime \prime }}{r}$,
and we have
\begin{eqnarray}
\eta =-\frac{M+2}{rM}\overset{\frac{N}{2}}{%
\underset{j=\left\langle \left\langle \frac{N}{2}\right\rangle
\right\rangle
}{\sum }}\frac{d_{j}}{\left( j+1\right) }\overset{j}{\underset{m=-j}{\sum }}%
m\left( \frac{1+r}{2}\right) ^{\frac{N}{2}-m}\left( \frac{1-r}{2}\right) ^{%
\frac{N}{2}+m},
\end{eqnarray}
which depends on $r$, i.e., a parameter depending on the input state
$\rho $. When $r=1$, the input state is pure and the shrinking
factor reaches the optimal bound $\eta =\frac{N\left( M+2\right)
}{M\left( N+2\right) }$. While when $r\neq 1$, we find $ \eta
>\frac{N\left( M+2\right) }{M\left( N+2\right) }$, that is to say,
for mixed state broadcasting the shrinking factor can larger than
that of optimal pure state cloning, though which {\it does} depend
on the purity of input mixed states.

\section{Conclusion}

We introduce a pure states decomposition for $N$ identical mixed
states $\rho ^{\otimes N}$. An optimal quantum broadcasting is
proposed to copy the mixed states input. We show that the mixed
states can be quantum broadcasted equally well as that of the pure
states in the sense that the shrinking factors are the same for both
cases. The
shrinking factor is a constant and is thus independent of each input qubit $%
\rho $, and in this sense we say it is universal. This broadcasting
procedure is optimal since the shrinking factor is optimal. The
optimal broadcasting procession for mixed states is not unique, and
we also present other different quantum broadcasting procedures with
similar properties, i.e.,universal and optimal.

In the broadcasting procession, we assume that each input qubit is
an arbitrary mixed state $\rho $, no prior information is available.
In case that partial information of the initial state is known, it
is generally expected that a higher shrinking factor can be
achieved. Such as the super-broadcasting scheme
\cite{Cirac,Giacomo,Buscemi}, a higher shrinking factor can be
achieved which depends on the purity of the mixed input state. For
the cloning scheme proposed in this paper, we restrict that the
input state can be arbitrary, pure or mixed, and no prior
information is available. Still we let the shrinking factor be {\it
universal}. Thus our quantum broadcasting provides a unified scheme
to broadcast mixed states and pure states.

\textit{Acknowlegements}: The authors are supported in part by `Bairen'
program, '973' program (2006CB921107), and NSFC. Also, this work was supported
in part by the Project of Knowlege Innovation Program (PKIP) of Chinese Academy of
Sciences.

\end{document}